\documentclass[preprint]{aastex}
\author{
Anastasia~Tsvetkova\altaffilmark{{1,2,*}}, 
Luciano Burderi\altaffilmark{1,3,4}, 
Alessandro Riggio\altaffilmark{1,3,5}, 
Andrea Sanna\altaffilmark{1,3,4},
Tiziana Di Salvo\altaffilmark{6}, 
}
\altaffiltext{1}{Dipartimento di Fisica, Universit\'{a} degli Studi di Cagliari, SP Monserrato-Sestu, km 0.7, I-09042~Monserrato, Italy}

\altaffiltext{2}{Ioffe Institute, Politekhnicheskaya 26, St.~Petersburg 194021, Russia}

\altaffiltext{3}{INFN, Sezione di Cagliari, Cittadella Universitaria, 09042 Monserrato, CA, Italy}

\altaffiltext{4}{INAF---Osservatorio Astronomico di Cagliari, Via della Scienza 5, 09047 Selargius (CA), Italy}

\altaffiltext{5}{INAF---Istituto di Astrofisica Spaziale e Fisica Cosmica di Palermo, Via U. La Malfa 153, 90146 Palermo, Italy}

\altaffiltext{6}{Dipartimento di Fisica e Chimica, Universit\'{a} degli Studi di Palermo, Via Archirafi 36, 90123 Palermo, Italy}

\altaffiltext{*}{tsvetkova.lea@gmail.com}
\usepackage{cmap} 
\usepackage[english]{babel}
\usepackage{indentfirst}
\usepackage{amssymb}
\usepackage[unicode=true,draft]{hyperref}
\usepackage{pdflscape,natbib,color}
\usepackage{graphicx}
\usepackage{gensymb}
\usepackage{amsmath}
\usepackage{hyperref}
\slugcomment{\small{Accepted by MDPI Universe}}

\title{A Concept of Assessment of LIV Tests with \textit{THESEUS} Using the Gamma-Ray Bursts Detected by \textit{Fermi}/GBM}

\begin{document}
\newcounter{cit}
\newcommand{\cititem}[1]{\refstepcounter{cit}(\arabic{cit})~\label{Gen:#1}\citealt{#1}}
\newcommand{\KW}{Konus-\textit{Wind} }
\newcommand{\BAT}{\textit{Swift}/BAT }

\begin{abstract}

According to Einstein's special relativity theory, the speed of light in a vacuum is constant for all observers. 
However, quantum gravity effects could introduce its dispersion depending on the energy of photons.
The investigation of the spectral lags between the gamma-ray burst (GRB) light curves recorded in distinct energy ranges could shed light on this phenomenon:
the lags could reflect the variation of the speed of light if it is linearly dependent on the photon energy and a function of the GRB redshift.
We propose a methodology to start investigating the dispersion law of light propagation in a vacuum using GRB light curves.
This technique is intended to be fully {exploited using} the GRB data collected with \textit{THESEUS}.

\end{abstract}

\section{Introduction}

According to Einstein's special relativity theory, a proper length is Lorentz-contracted by a factor of $\gamma^{-1} = [1 - (v/c)^2]^{1/2}$ as observed from the reference frame moving at speed $v$ relative to the rest frame.
However, various spacetime theories, e.g., some string or loop quantum gravity theories (see, e.g., \cite{Rovelli1988, Rovelli1990, Rovelli1998}), imply the existence of a minimum spatial length of the order of the Plank length $l_\textrm{Pl} = \sqrt{G \hbar /c^3} = 1.6 \times 10^{-33}$~cm~\cite{Hossenfelder2013}.
Lorentz invariance is a fundamental property of both the standard model of particle physics and general relativity.
In general relativity, a locally inertial reference frame where the Lorentz symmetry is fulfilled can always be chosen.
Some quantum gravity (QG) theories predict the Lorentz invariance violation (LIV) at the Planck {energy} scale ($E_\textrm{Pl} = \sqrt{\hbar c^5/G} \simeq 1.22 \times 10^{19}$~GeV) as there exists a minimum spatial length $l_\textrm{min} = \alpha l_\textrm{Pl}$ (where $\alpha \sim  1$ is a dimensionless constant inherent to a particular spacetime theory), which, e.g., in the string theories, corresponds to the string length.
Therefore, the Lorentz contraction is limited by this spatial scale (see \cite{Hossenfelder2013} for a review).

There are several frameworks implying LIV, e.g., string theory \cite{Kostelecky1989a, Kostelecky1989b}, noncommutative spacetime \cite{Carroll2001, Ferrari2007}, Brane worlds \cite{Santos2013}, Ho\v{r}ava--Lifshitz gravity \cite{Horava2009}.
LIV was considered in the gravitational context for the first time in \cite{Kostelecky2004}, where the so-called standard model extension was developed.
The Bumblebee\footnote{The name of the model was inspired by the insect, whose ability to fly has been questioned theoretically.} models, which are the simplest cases of theories including the spontaneous breaking of Lorentz symmetry, are effective field QG theories describing a vector field with a non-zero vacuum expectation value and involving the vacuum condensate \cite{Kostelecky1989a, Kostelecky1989b}.
The spontaneous symmetry breaking preserves both the geometric constraints and conservation laws or quantities required by the general relativity theory or Riemannian geometry.
Regarding gravity, LIV can happen if a vector field ruled by a potential exhibiting a minimum rolls to its vacuum expectation value, similar to the Higgs mechanism \cite{Kostelecky2004}. 
This ``bumblebee'' vector thus takes an explicit (four-dimensional) orientation, and preferred-frame effects may emerge \cite{Bertolami2005}.
{The generalized uncertainty principle (GUP) states that, in quantum theory, if the quantities are incompatible, they are mutually dependent, and measuring one observable may yield some information about its incompatible partners.}
GUP is based on a momentum-dependent modification of the standard dispersion relation, which is supposed to produce LIV \cite{Tawfik2016, Lambiase2018}.
See, e.g.,\cite{Kanzi2019, Ovgun2019, Kanzi2021, Delhom2021, Gogoi2022, Gogoi2023, Neves2023} for recent advances in the Bumblebee and GUP models.

One of the LIV effects, relevant to astrophysics, is the existence of a dispersion law for the photon speed $c$ (see, e.g., \cite{Amelino-Camelia2000}).
However, LIV is not a mandatory property of all QG theories: in some of them, e.g., in the spacetime uncertainty principle \cite{Burderi2016} or in the quantum spacetime \cite{Sanchez2019}, LIV is not expected.
Despite photon velocity dispersion, the Lorentz invariance is not violated and the dispersion law is a second-order effect relative to the ratio of the photon energy to the QG energy scale.
To obtain an idea of the nature of the speed of light dispersion due to LIV, e.g., within the Liouville string approach, one can consider a vacuum as a non-trivial medium containing ``foamy'' quantum gravity fluctuations whose origin can be imagined as processes that involve the pair creation of virtual black holes.
In this concept, one can verify that the massless particles of different energies can excite vacuum fluctuations differently as they propagate through the quantum gravity medium, producing  a non-trivial dispersion relation of Lorentz ``non-covariant'' form, similarly to the thermal medium \cite{Amelino-Camelia1998}. For more details regarding this concept, see, e.g., \cite{Amelino-Camelia1997}.

Since gamma-ray bursts (GRBs) are characterized by high-energy emission, large cosmological distances, and temporal variability at short ($\lesssim$10~ms) timescales, they have been applied as powerful tools in LIV searches (see, e.g., \cite{Amelino-Camelia1998, Ellis2003, Ellis2006, Abdo2009a, Vasileiou2015, Zhang2015, Pan2015, Xu2016, Xu2016a, Chang2016, Wei2017, Ganguly2017, Liu2018, Zou2018, Ellis2019, Wei2019, Pan2020, Acciari2020, Du2021, Agrawal2021, Wei2021, Bartlett2021, Xiao2022, Desai2023}) for more than two decades.
{In \cite{Amelino-Camelia1998, Amelino-Camelia1998err}}, {it was} first suggested to test LIV using a comparison of the arrival times of GRB photons detected in distinct energy ranges.
In \cite{Abdo2009a, Vasileiou2013, Vasileiou2015}, the authors exploited the spectral lag between high-energy ($\sim$31~GeV) and low-energy photons of GRB~090510 to derive lower limits on the linear and quadratic QG energy: $E_\textrm{QG,1} > (1-10) \times E_\textrm{Pl} 
 = 1.22 \times (10^{19} - 10^{20})$~GeV and $E_\textrm{QG,2} > (1-10) \times E_\textrm{Pl} = 1.3 \times 10^{11}$~GeV, respectively.
{In \cite{Abdo2009b}, $E_\textrm{QG} > 10^{18}$~GeV was obtained} based on the observation of GRB~080916C.
For GRB~190114C \cite{MAGIC2019}, the linear and quadratic LIV were constrained using the time delays of TeV photons: $E_\textrm{QG,1} > 0.58 \times 10^{19}$~GeV ($E_\textrm{QG,1} > 0.55 \times 10^{19}$~GeV) and $E_\textrm{QG,2} > 0.63 \times 10^{11}$~GeV ($E_\textrm{QG,1} > 0.56 \times 10^{11}$~GeV) for the subluminal (superluminal) case \cite{Acciari2020}.
{In \cite{Liu2022}, the authors} constrained the linear and quadratic LIV for the set of 32~\textit{Fermi}/GBM GRBs with known redshifts and characterized by the positive-to-negative transition of the spectral lag: $E_\textrm{QG,1} = 1.5 \times 10^{14}$~GeV for the linear case and $E_\textrm{QG,2}~=~8~\times~10^{5}$~GeV for the quadratic case.

The phenomenon of GRBs remains puzzling, although much progress has been made. 
Both the light curves and the spectra among GRBs vary significantly.
It is generally believed that collisions between relativistic shells ejected from an active central engine produce pulses in GRB light curves \cite{Rees1994}.
The collision of the slower-moving shell with the second, faster shell ejected later produces a shock that dissipates internal energy and accelerates the particles that emit the GRB radiation. 
As of now, two ``physical'' classes of GRBs are distinguished (see, e.g., \cite{Zhang2009}): the merger-origin Type~I GRBs \cite{Blinnikov1984, Paczynski1986, Eichler1989, Paczynski1991}, which are usually short, with a duration of less than 2~s \cite{Mazets1981, Kouveliotou1993}, and spectrally hard, and the collapser-origin Type II GRBs \cite{Woosley1993, Paczynski1998, MacFadyen1999, Woosley2006}, characterized by longer durations.

The spectral lag, which is known as the difference in arrival time between high-energy and low-energy photons, is a common phenomenon occurring during the GRB prompt emission phase \cite{Norris1986, Norris2000, Band1997, Chen2005} and in high-energy astrophysics in general (e.g., \cite{Norris2000, Zhang2002}).
{The authors of \cite{Cheng1995}} were the first to analyze the spectral lags of GRBs.
It was found that a soft lag, i.e., the hard photons arriving first, dominates in long GRBs \cite{Norris2000, Wu2000, Chen2005, Norris2005}, and some GRBs have significantly different spectral lags in early and late epochs \cite{Hakkila2004}.
It was shown that the lags are correlated with the GRB luminosity \cite{Norris2000} and the jet break times in afterglow light curves \cite{Salmonson2002}; the spectral lags of long GRBs are correlated with the pulse duration, while the spectral lags of short GRBs are not \cite{Yi2006}.
The GRB spectral lags can be explained within several physical models, such as the curvature effect of a relativistic jet and rapidly expanding spherical shell \cite{Ioka2001, Shen2005, Lu2006, Shenoy2013, Uhm2016}. 
Regardless of its physical origin, a spectral lag is an important GRB parameter as it may help to distinguish between long and short GRBs: long bursts have large lags, while short bursts have relatively smaller or negligible lags \cite{Norris1995, Ryde2005, Norris2006}. 

To show the feasibility of the search for the dispersion of the speed of the light in the data collected by THESEUS \cite{Amati2018}, one can {start performing} similar research using the data of already commissioned missions.
In this paper, we describe a methodology for the testing of the dispersion of the light speed, {and we intend to apply it to} the \textit{Fermi}/GBM data and the set of GRBs with known redshifts.
The paper is organized as follows.
We start with a brief description of the instrumentation and data in Section~\ref{sec:data}. 
The methodology that involves constraining LIV using GRBs is described in Section~\ref{sec:methods}. 
In Section~\ref{sec:discussion}, we discuss the \textit{THESEUS} mission in the context of the LIV tests using GRBs.
Section~\ref{sec:conclusion} concludes the paper.

\section{Instrumentation and Data}
\label{sec:data}
Among several space-based detectors able to collect GRB data, the most prolific are \textit{Swift}/BAT (BAT; \cite{Gehrels2004}), \textit{Fermi}/GBM (GBM; \cite{Meegan2009}), and Konus-\textit{Wind} (KW; \cite{Aptekar1995}).
More details regarding the BAT, GBM, KW, and other GRB detectors' design and performance can be found in~\cite{Tsvetkova2022}.
However, since \textit{Swift}/BAT collects GRB data in a narrow energy band of 15--350~keV, and Konus-\textit{Wind} records GRB light curves in three fixed energy windows, the \textit{Fermi/GBM} GRB time histories seem to be the most suitable for the testing of the speed of light variance.

Launched in June~2008, the \textit{Fermi} Gamma-Ray Space Telescope \cite{Thompson2022} harbors two scientific instruments: the Gamma-Ray Burst Monitor (GBM) and the Large Area Telescope (LAT; \cite{Atwood2009}).
{The LAT covers the 30~MeV--300~GeV band, while the GBM, intended to detect and study GRBs, is sensitive within the 8~keV--30~MeV energy range, extending the spectral band over which bursts are observed downwards to the hard X-ray range.
}GBM comprises twelve NaI(Tl) detectors covering an energy range of 8~keV$^{-1}$~MeV and two bismuth-germanate (BGO) scintillation detectors sensitive within the 150~keV to 30~MeV band that observe the whole sky not occluded by the Earth (>8~sr). 

The primary scientific data produced by GBMs can be summarized as a time history {and} spectra, which {are} provided as temporally pre-binned (CTIME and CSPEC) or temporally unbinned time tagged events (TTE). 
These data types are produced as ``snippets'' for every trigger and are also provided continuously. 
The CTIME data are collected in 8~energy channels with a 256~ms time resolution, while the CSPEC data are recorded in 128~channels with an 8.192~s time resolution.
TTEs for each detector are recorded with time precision down to 2 $\mu$s, in 128 energy channels, matching the CSPEC ones, which gives an excellent opportunity to bin the data in time and energy in a suitable way.
From 2008 through November 2012, TTE were available only during a 330~s interval: from 30~s before the burst trigger to 300~s after the burst trigger. 
Since November 2012, GBM flight software has produced a new data type, continuous TTE (CTTE), available at all times that the instrument is~operating.

To date, the GBM has triggered almost 3500 GRBs, among which almost 3000 have $T_{90} > 2$~s. 

\section{Methodology}
\label{sec:methods}
We start the research by computing the redshifts and the rest-frame spectral lags for all GBM-triggered GRBs.
Then, thanks to the central limit theorem, we can consider the distribution of the lags as normal and compute the mean and variance.
The large number of ``measurements'' is expected to significantly increase the accuracy of the spectral lags and constraints on $E_\textrm{QG}$.
Another approach would be to constrain the QG energy (see Section~\ref{sec:lags}) first and then estimate its mean values and variance.

\subsection{Brief Introduction to the Basics of the Speed of Light Variance}
\label{sec:lags}
The velocity dispersion law for photons can be expressed as a function of its observer-frame energy $E_\textrm{obs}$ in units of the QG energy scale $E_\textrm{QG}$, at which the quantum nature of gravity becomes important:
\begin{equation}  
\label{eq:v2c}
v_\textrm{phot}/c - 1 = \xi \left(\frac{E_\textrm{obs}}{E_\textrm{QG}} \right)^n,
\end{equation}
where $v_\textrm{phot}$ is the group velocity of a photon wave-packet, 
$E_\textrm{QG} = \zeta m_\textrm{Pl} c^2 = \zeta E_\textrm{Pl}$, \linebreak{$m_\textrm{Pl}  = 2.176 \times 10^{-5}$~g is the Planck mass},
$\zeta \sim \alpha^{-1} \sim 1$ expresses the significance of the QG effects,
$\xi \sim \pm 1$ is a dimensionless constant inherent to a particular QG theory, 
and the index $n$ denotes the order of the first relevant term of the small parameter $\left(\frac{E_\textrm{obs}}{E_\textrm{QG}} \right)$.
This expression takes into account that high-energy photons can travel faster (superluminal, $\xi = +1$) or slower (subluminal, $\xi = -1$) than low-energy ones \cite{Amelino-Camelia2009}.  

The difference in the arrival times of photons emitted at the same time in the same place is
\begin{equation}
\label{eq:dt}    
\Delta t_\textrm{QG} = \xi \left( \frac{D_\textrm{trav}}{c}\right) \left( \frac{\Delta E_\textrm{obs}}{E_\textrm{QG}} \right)^n,
\end{equation}
where $D_\textrm{trav}$ is the comoving distance traversed by a massless particle, emitted at redshift $z$ and traveling down to redshift 0.

\subsection{Observed Spectral Lag as a Function of Redshift}
\label{lags}
For a given observer-frame energy $E_{\rm obs}$, the total spectral lag $\tau_{\rm total, \, obs}(E_{\rm obs}, z)$
can be split into two terms: 
\begin{equation}
\label{eq:tauobs}
\tau_{\rm total, \, obs}(E_{\rm obs}, z) = \tau_{\rm int, \, obs}(E_{\rm obs}, z) + \tau_{\rm QG, \, obs}(E_{\rm obs}, z),
\end{equation}
where $\tau_{\rm int, \, obs}$ is the observed intrinsic spectral lag, which corresponds to the intrinsic rest-frame lag 
\begin{equation}
\label{eq:tauintrf}
\tau_{\rm int, \, rf}(E_\textrm{obs}) = \tau_{\rm int, \, obs}(E_{\rm rf})/(1 + z)
\end{equation}
induced by the GRB central engine emission mechanism and assumed to be independent of the  {photon source} redshift $z$. $\tau_{\rm QG, \, obs}$ is the lag induced by the QG effects discussed above, and {and $E_\textrm{rf}$ is the photon energy in the rest frame of its source.}
Following \cite{Jacob2008}, $\tau_{\rm QG, \, obs}$ can be expressed as a function of the GRB rest-frame energy:
\begin{equation}
\label{eq:tauqgobs}
\begin{split}
\tau_{\rm QG}(E_{\rm rf}, z) = \xi\left(\frac{1}{H_{0}} \right)
\left( \frac{E_{\rm rf}}{\zeta E_{\rm pl}} \right)^n
\left( \frac{1+n}{2} \right) \left( \frac{1}{1+z} \right)^n \times \\
\times \int_0^z \frac{(1+z')^n dz'}{\sqrt{\Omega_{m}(1+z')^3 + \Omega_{\Lambda}}}{,}
\end{split}
\end{equation} 
{where $H_0$ is the Hubble constant, $\Omega_{m}$ is the matter density parameter, and $\Omega_{\Lambda}$ is the dark energy density parameter, i.e., the parameters of the standard $\Lambda$CDM model.}

Experimentally, the total observed spectral lag $\tau_{\rm total, \, obs}(E_{\rm rf}, z)$ {can be} computed by cross-correlating the GRB light curves recorded in the redshift-{dependent} energy windows corresponding to the fixed rest-frame energy windows as $E_{\rm obs} = E_{\rm rf} / (1+z)$, with the GRB {light curve} collected at the lowest possible energy channel, where the lag induced by QG is negligible, e.g., it is $\sim$$\mu$s in the  5--20  keV energy range, while, in the higher-energy bands, it is $\sim$ms. 

The total observed spectral lag $\tau_{\rm total, \, obs}(E_{\rm rf}, z)$  should follow the relation obtained by inserting Equations (\ref{eq:tauintrf})  and (\ref{eq:tauqgobs})  into Equation (\ref{eq:tauobs}):
\begin{equation}
\label{eq:taurfobs}
\begin{split}
\tau_{\rm total, \, obs}(E_{\rm rf}, z) =
\tau_{\rm int, \, obs}(E_{\rm rf}) + 
\\
\xi\left( \frac{1}{H_{0}} \right)
\left( \frac{E_{\rm rf}}{\zeta E_{\rm pl}} \right)^n 
\left( \frac{1+n}{2} \right)
\left( \frac{1}{1+z} \right)^n
\int_0^z \frac{(1+z')^n dz'}{\sqrt{\Omega_{m}(1+z')^3 + \Omega_{\Lambda}}}.
\end{split}
\end{equation}

The transformation of spectral lags from the observer frame back to the rest frame\footnote{This value reflects only the mathematical transformation from the observer frame to the rest frame as, at the moment of emission of two photons, the QG lag between them equals zero.} can be performed by dividing by the redshift factor $(1+z)$:
\begin{equation}
\tau_{\rm total, \, rf}(E_{\rm rf}, z) = \tau_{\rm total, \, obs}(E_{\rm rf}, z) / (1+z).
\end{equation}

Thus, the GRB rest-frame spectral lag obeys the following relation:
\begin{equation}
\label{eq:thetaobs}
\begin{split}
\tau_{\rm total, \, rf}(E_{\rm rf}, z) =
\frac{\tau_{\rm int, \, obs}(E_{\rm rf})}{(1+z)} + \\ 
\xi\left( \frac{1}{H_{0}} \right)
\left( \frac{E_{\rm rf}}{\zeta E_{\rm pl}} \right)^n 
\left( \frac{1+n}{2} \right)
\left( \frac{1}{1+z} \right)^{n+1}
\int_0^z \frac{(1+z')^n dz'}{\sqrt{\Omega_{m}(1+z')^3 + \Omega_{\Lambda}}}.
\end{split}
\end{equation}

Let us define a function
\begin{equation}
\label{eq:uzeta}
\begin{split}
u(z) =  
\left( \frac{1+n}{2} \right)
\left( \frac{1}{1+z} \right)^{n+1}
\int_0^z \frac{(1+z')^n dz'}{\sqrt{\Omega_{m}(1+z')^3 + \Omega_{\Lambda}}}.
\end{split}
\end{equation}

Then, the experimentally determined lag $\tau_{\rm total, \, obs}(E_{\rm rf}, z)$ will follow the relation
\begin{equation}
\label{eq:thetau}
\begin{split}
\tau_{\rm total,rf}(E_{\rm rf}, z) =
\tau_\textrm{int,rf}(E_{\rm rf}) + \xi\left( \frac{1}{H_{0}} \right)
\left( \frac{E_{\rm rf}}{\zeta E_{\rm pl}} \right)^n
u(z).
\end{split}
\end{equation}

The dependence of $\tau_{\rm total, \,rf}(E_{\rm rf}, z)$ on $u(z)$ is expected to be linear, with the intercept corresponding to the intrinsic lag and the slope proportional to 
the ratio of the rest-frame photon energy to the QG energy $\zeta E_{\rm pl}$
raised to the $n$-th power, which is the first significant term in the series expansion of the quantum gravity dispersion relation.
An argument in support of the independence of $\tau_{\rm int, \, rf}$ from $z$ is the absence of the prominent cosmological evolution of GRB energetics \cite{Tsvetkova2017, Tsvetkova2022}, which indicates that the GRB central engine does not evolve significantly with $z$. 
However, if $\tau_{\rm int, \, rf}$ is {dependent} on $z$, one can fit both $\tau_{\rm int, \, rf}(z)$ and $\tau_\textrm{QG}(z)$ simultaneously to the data, as was done in \cite{Liu2022}.

\begin{figure}[]
\includegraphics[width=0.8\textwidth]{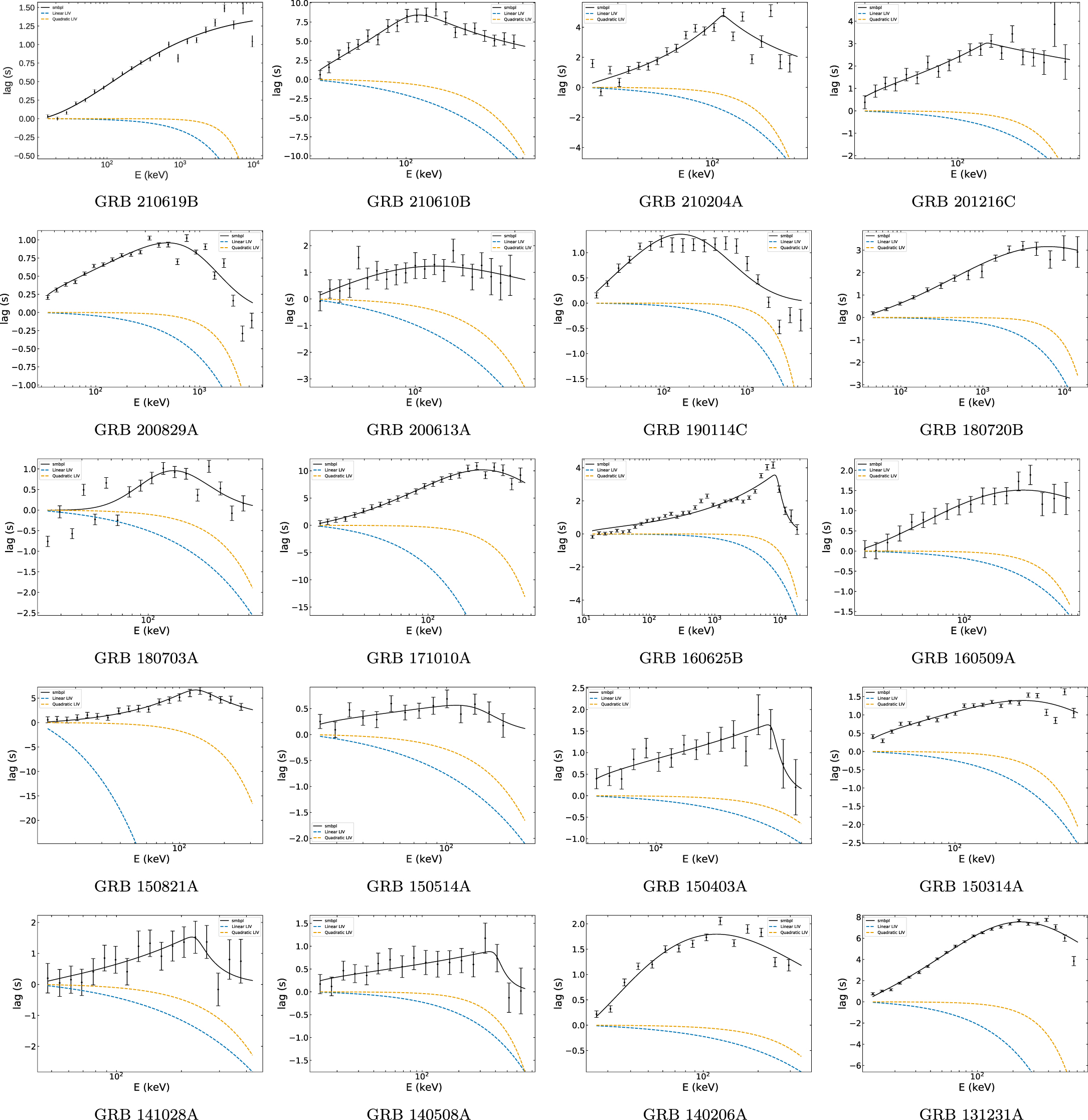}
\caption{
{The dependence of the spectral lag on the energy window for the GRB sample studied in~\cite{Liu2022}. 
The fits with a smoothly broken power law (SBPL) are shown by black solid lines. 
Blue and orange dotted lines denote the maximally allowed LIV-induced lags in linear and quadratic cases, thereby defining the lower limits on the QG energy.
{This figure is adapted from Figure~1 from~\cite{Liu2022} (see Section~\ref{sec:methods} of this work for details).}}
\textcopyright AAS. Reproduced with permission.}
\label{fig:lag-energy}
\end{figure}

{In \cite{Liu2022}, it was} found that the behavior of the lags follows some key statistical properties described in terms of log-normal and Gaussian distributions around average values. 
These empirical functions describing the spectral lags (as a function of energy) for each of the 32 \textit{Fermi}/GBM GRBs in the sample {of bursts with known redshifts that exhibit the lag transition phenomenon} are shown in Figure~\ref{fig:lag-energy} (as derived from Figure~1 of~\cite{Liu2022}).
In the method proposed here, we make the reasonable assumption that the intrinsic lags (that dominate in magnitude over the small delays induced by QG effects; see the upper limits for the first- and second-order QG delays shown as blue and orange curves in Figure~\ref{fig:lag-energy}) do not correlate with the redshift as the distance of a given GRB from us does not affect the emission properties in, {e.g.,} the fireball model.
This means that, averaging over a large sample of GRBs at different redshifts, the intrinsic delays will cluster around the common value that defines the intrinsic average rest-frame lag. 

\subsection{The Technique of Averaging over the Sample}
\label{technique}
Relation (\ref{eq:thetau}) shows that, for a given GRB rest-frame energy $E_{\rm rf}$, the lag $\tau_{\rm total, \, rf}(E_{\rm rf})$ depends only on the GRB redshift $z$ through the function 
$u(z)$ defined in Equation (\ref{eq:uzeta}). 
Therefore, fixing $E_{\rm rf}$ for an ensemble of $N$ GRBs with known redshift, one can compute the function $\tau_{\rm total, \, rf}(E_{\rm rf})$ from the {observed} data for all GRBs of the ensemble, 
i.e., it is possible to obtain a set of $N$ experimentally computed values of $\tau_{\rm total, \, rf}(E_{\rm rf})$. 
This ensemble of $N$ values can be fitted as a function of $u(z)$ through Equation (\ref{eq:thetau}) to obtain the best fit values of the intrinsic lag in the GRB rest frame $\tau_{\rm int, \, rf}$, and the coefficient of the QG-induced delay at the GRB rest-frame energy $E_{\rm rf}$
\begin{equation}
\label{eq:bestfit}
\begin{split}
\left\{
\begin{array}{lclcl}
\left[ \tau_{int, \, rf}(E_{\rm rf}) \right]_{\rm BEST} & &\vspace{6pt} \\ \vspace{6pt}
{\rm and}  & & & &\\\vspace{6pt}
\left[ \xi\left( \frac{1}{H_{0}} \right) \left( \frac{E_{\rm rf}}{\zeta E_{\rm pl}} \right)^n \right]_{\rm BEST} & = & & & \\\vspace{6pt}
\left[ \xi\left(\frac{1}{H_{0}} \right) \left( \alpha\frac{E_{\rm rf}}{ E_{\rm pl}} \right)^n\right]_{\rm BEST} & = & 
\left[ \phi(\alpha
 E_{\rm rf}) \right]_{\rm BEST} & & \\
\end{array}
\right. .
\end{split} 
\end{equation}

Once the values of $\left[ \phi(\alpha E_{\rm rf}) \right]_{\rm BEST}$ are obtained for all $ E_{\rm rf}$, 
these values can be
 plotted as a function of $s(E_{\rm rf}) = (E_{\rm rf}/E_{\rm pl})^n$ and subsequently linearly 
fitted through the equation\vspace{6pt}
\begin{equation}
\label{eq:esse}
\begin{split}
\phi(\alpha E_{\rm rf}) = \left(\frac{\alpha
^n}{H_{0}} \right) s(E_{\rm rf}) = \Delta_{\rm QG} \, s(E_{\rm rf})
\end{split},
\end{equation}
to obtain the best fit value of the strength of the QG effect 
$\Delta_{\rm QG} = \alpha^n/H_{0}$.
We note that this technique allows us to combine the whole ensemble of $N$ GRBs to obtain a unique measure
of the strength of the QG effect, whose uncertainty $\sigma_{\Delta_{\rm QG}}$, in the absence of other systematic errors, 
only depends on the (Poissonian) statistics of the whole ensemble of $N$ GRBs and therefore improves as the 
inverse square root of $N$:
\begin{equation}
\label{eq:zeta}
\begin{split}
\sigma_{\Delta_{\rm QG}} \propto \frac{1}{\sqrt{N}}.
\end{split}
\end{equation}

Consequently, the precision of the measurement of the QG effect's strength can be improved by increasing the size of the {analyzed} sample.

\subsection{GRB Intrinsic Spectral Lags vs. Quantum Gravity Effects}
\label{sec:intrinsic_vs_qg}
The GRB spectral lag can be caused by a mixture of two effects: the QG one and the one inherent to the fireball model.
The latter is due to the curvature effect, i.e., the kinematic effect caused by the fact that the observer looks at an increasingly off-axis annulus area relative to the line-of-sight \cite{Fenimore1996, Salmonson2000, Kumar2000, Ioka2001, Qin2002, Qin2004, Dermer2004, Shen2005, Lu2006}. 
Softer low-energy radiation comes from the off-axis annulus area with smaller Doppler factors and is delayed for the observer with respect to on-axis emission due to the geometric curvature of the shell.

A competing hypothesis is that the traditional view {based on} the high-latitude emission ``curvature effect'' of a relativistic jet cannot {explain} spectral lags.
{Instead, spectral peaks should be swept across the observing energy range in a specific manner to account for the observed spectral lags.}
{A simple physical model {that implies} synchrotron radiation from a rapidly expanding outflow can explain GRB spectral lags \cite{Uhm2016}. 
This model requires the following conditions to be fulfilled: 
(1) the emission radius has to be large (over several $10^{14}$~cm from the central engine), in the optically thin region, well above the photosphere;
(2) the $\gamma$-ray photon spectrum is curved (as observed);
(3) the magnetic field strength in the emitting region decreases with the radius as the region expands in space, which is consistent with an expanding jet;
and (4) the emission region itself undergoes rapid bulk acceleration as the prompt emission is produced.
These requirements are consistent with a Poynting-flux-dominated jet abruptly dissipating magnetic energy at a large distance from the engine.}
The aforementioned theories successfully explain the positive spectral lags.
Nevertheless, the rarely observed negative lags remain a more
intriguing phenomenon that can be used to infer the different radiation mechanisms \cite{Li2010, Zhang2011} or emission regions \cite{Toma2009} of low- and high-energy photons.

For a given GRB, the intrinsic delay inherent to the GRB emission could mimic the genuine quantum gravity eﬀect, making these two effects difficult to disentangle. 
However, currently, there is no evidence for a correlation between the GRB intrinsic delays and the distances to its sources.
For example, in \cite{Tsvetkova2017, Tsvetkova2021}, where the largest sample of GRBs with known redshifts detected by a single instrument in a wide energy range is studied, the significance of the cosmological evolution of GRB energetics is $\lesssim$2$\sigma$.
Meanwhile, the delays induced by a photon dispersion law are proportional both to the light travel distance (a function of redshift) and to the diﬀerences in the energy of the photons. 
This {dual} dependence on energy and redshift {could be} the unique {feature} of a genuine QG eﬀect. 
As suggested in \cite{Burderi2020}, given an adequate collection area, GRBs, once their redshifts are known, are potentially excellent tools to search for the ﬁrst-order dispersion law for photons.

\subsection{Computation of Spectral Lags}
\label{sec:comp_lag}
\begin{figure}[]
\includegraphics[width=0.8\textwidth]{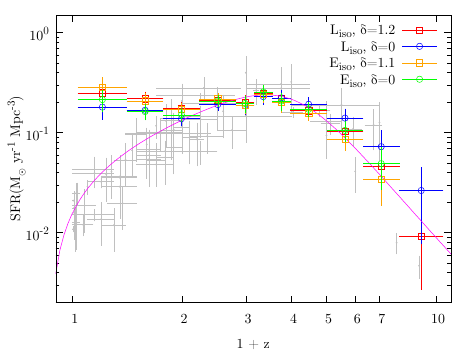}
\caption{{The cosmological GRB formation rate (GRBFR) derived in \cite{Tsvetkova2021}, superposed onto the star formation rate (SFR) data from the literature.}
The gray points show the SFR data from \cite{Hopkins2004, Bouwens2011, Hanish2006, Thompson2006}. 
The {marked} line denotes the SFR approximation from \cite{Li2008}. 
{The GRBFR normalization is equal for all four data sets} and the GRBFR points have been shifted arbitrarily to match the SFR at $(1 - z) \sim 3.5$.
Figure 5b from \cite{Tsvetkova2021}. \textcopyright AAS. Reproduced with permission.}
\label{fig:grbfr}
\end{figure}

To avoid {distortions} due to the fact that the shape of the light curve changes with the energy, we suggest fixing the energy channels in which the light curves are recorded to certain values in the rest frame. 
In this case, the corresponding observer-frame values of the channel boundaries will be $E_\textrm{obs} = E_\textrm{rest}/(1+z)$, i.e., redshift-{dependent}. 
{The first step to test the LIV effects with the suggested technique would be apply it to the GBM data. 
Thus, we propose to use the following energy bands to record the light curves.}
Given the 9~keV lower boundary of the GBM spectral window, one has to select the rest-frame channels starting from, at least, 60~keV, to allow for bursts with redshifts up to $z = 5$.
This boundary can be shifted towards a higher value to allow high-redshift GRBs to contribute to the study.
However, we should mention that the majority of GRBs have redshifts $z < 5$ (see Figure~\ref{fig:grbfr}).
Some examples of the pseudo-logarithmic channels that could be used are 60--100~keV, 100--160~keV, 160--250~keV, 250--400~keV, 400--600~keV, 600--900~keV, or 60--80~keV, 80--100~keV, 100--130~keV, 130--160~keV, 160--200~keV, 200--250~keV, 250--320~keV, 320--400~keV, 400--500~keV, 500--650~keV, 650--900~keV.
We found these numbers of channels to be reasonable in terms of SNR based on the research of \cite{Liu2022}, carried out for 32 \textit{Fermi}/GBM GRBs.

Since the GRB energetics are usually considered on the logarithmic scale, we suggest adopting the geometric mean of the lower and upper boundaries of an energy band, $E_\textrm{phot} = \sqrt{E_\textrm{min} \times E_\textrm{max}}$, as a proxy for the average energy of photons in the given range.
\vspace{-3pt}

Since the spectral lag distributions of the short and long GRBs significantly differ \citep{Yi2006}, and these two types of bursts belong to distinct classes of progenitors, they have different intrinsic spectral lags. 
Thus, we suggest studying them separately.

\subsection{Obtaining GRB Redshifts}
{Since the suggested technique of testing LIV using GRBs strongly relies on prior knowledge of the burst redshift, it is necessary either to measure it directly from the observations in optics or estimate it using the prompt emission parameters.}
{GRB redshift measurements based on the detection of  emission lines or absorption features of GRB host galaxies imposed on the afterglow continuum, or performed photometrically, are widespread. }
{However, there are other methods to obtain the redshift estimates}, e.g., the ``pseudo-redshift'' (pseudo-z) technique based on the spectral properties of GRB prompt high-energy emission \cite{Atteia2003}, using well-known correlations such as, for example, the Norris correlation (spectral lag vs. isotropic peak luminosity; \cite{Norris2000}), the Amati correlation (rest-frame peak energy vs. isotropic energy release; \cite{Amati2002}), the isotropic peak luminosity vs. temporal variability correlation \cite{Reichart2001, Fenimore2000}, the Yonetoku (the rest-frame peak energy vs. the isotropic peak luminosity; \cite{Yonetoku2004}) correlation, etc., or the method of searching for a minimum on the intrinsic hydrogen column density versus the redshift plane (see, e.g., \cite{Ghisellini1999}).

Nowadays, the machine learning (ML) approach to redshift estimation is becoming popular in astrophysics (see, e.g., \cite{DIsanto2018, Dainotti2019, Lee2021, Momtaz2022}).
Supervised ML is a data mining method based on prior knowledge of a ``training'' data set, on which we can build models predicting the parameter under consideration, a ``validation'' set, which provides an unbiased evaluation of a model's fit while tuning the model's hyperparameters, and a ``test'' data set necessary for an unbiased evaluation of the final model fit. 

Considering only spectroscopic and photometric redshifts, there were $\gtrsim$420 GRBs with reliably measured redshifts by the middle of 2022 (for a list of GRBs with measured redshifts, see \cite{Gruber2011, Atteia2017, Tsvetkova2017, Minaev2020, Minaev2021_err, Tsvetkova2021}, the Gamma-Ray Burst Online Index\footnote{\url{https://sites.astro.caltech.edu/grbox/grbox.php}}, Jochen Greiner's GRB table\footnote{\url{https://www.mpe.mpg.de/~jcg/grbgen.html}}, and the references therein).
Using one of the aforementioned techniques, one can estimate the redshift of any burst based on its temporal or spectral parameters and energetics. 
For example, \cite{Lloyd-Ronning2002, Yonetoku2004, Kocevski2006} used various correlations to obtain unknown GRB redshifts from GRB observables, while \cite{Ukwatta2016} used the random forest algorithm to estimate GRB redshifts.

\section{Discussion}
\label{sec:discussion}
\textit{THESEUS} is a mission aimed at increasing the discovery rate of the high-energy transient phenomena over the entirety of cosmic history and fully exploiting GRBs to explore the early Universe \cite{Amati2018, Amati2021}.
\textit{THESEUS} is likely to become a cornerstone of multi-messenger and time-domain astrophysics thanks to its {exceptional} payload, providing wide and deep sky monitoring in a {broad} energy {range} (0.3~keV--20~MeV); focusing capabilities in the soft X-ray band, providing large grasp and a high angular resolution; and onboard near-IR capabilities for immediate transient identification and redshift determination.

The \textit{THESEUS} payload is planned to include the following instrumentation:
(1) the X-Gamma-Ray Imaging Spectrometer (XGIS, 2~keV--20~MeV): a set of two coded-mask cameras using monolithic X-gamma-ray detectors based on bars of silicon diodes coupled with a crystal scintillator, granting a $\sim$2~sr field of view (FoV) and source location accuracy of $\sim$10$^{\prime}$ in the 2--150~keV band, as well as a >4~sr FoV at energies $> 150$~keV, with a few $\mu$s timing resolution;
(2) a Soft X-Ray Imager (SXI, 0.3--5~keV): a set of two lobster-eye telescope units, covering a total FOV of $\sim$0.5~sr with source location accuracy $\lesssim 2^{\prime}$;
(3) an infrared telescope (IRT, 0.7--1.8~$\mu$m): a 0.7~m class IR telescope with a $15^{\prime} \times 15^{\prime}$ FOV, for a fast response, with both imaging (I, Z, Y, J, and H) and spectroscopic (resolving power, R$\sim$400, through $2^{\prime} \times 2^{\prime}$ grism) capabilities.

{Thanks to the unique combination of a wide 0.3~keV--10~MeV energy range, remarkable sensitivity, and exceptionally high counting statistics, \textit{THESEUS} heralds a new era in the multi-wavelength studies of GRBs, providing the community with a sample of GRBs with known redshifts of unprecedented size, which, in turn, will not only allow the use of GRBs as cosmological tools but also shed light on one of the most challenging aspects of QG theory, the systematic study of which is still beyond the current instrumental capabilities.
}
{The capability of \textit{THESEUS} to detect and localize GRBs, as well as measure their redshifts, will essentially surpass those of the current missions.}

\begin{figure}[]
\includegraphics[width=0.5\textwidth]{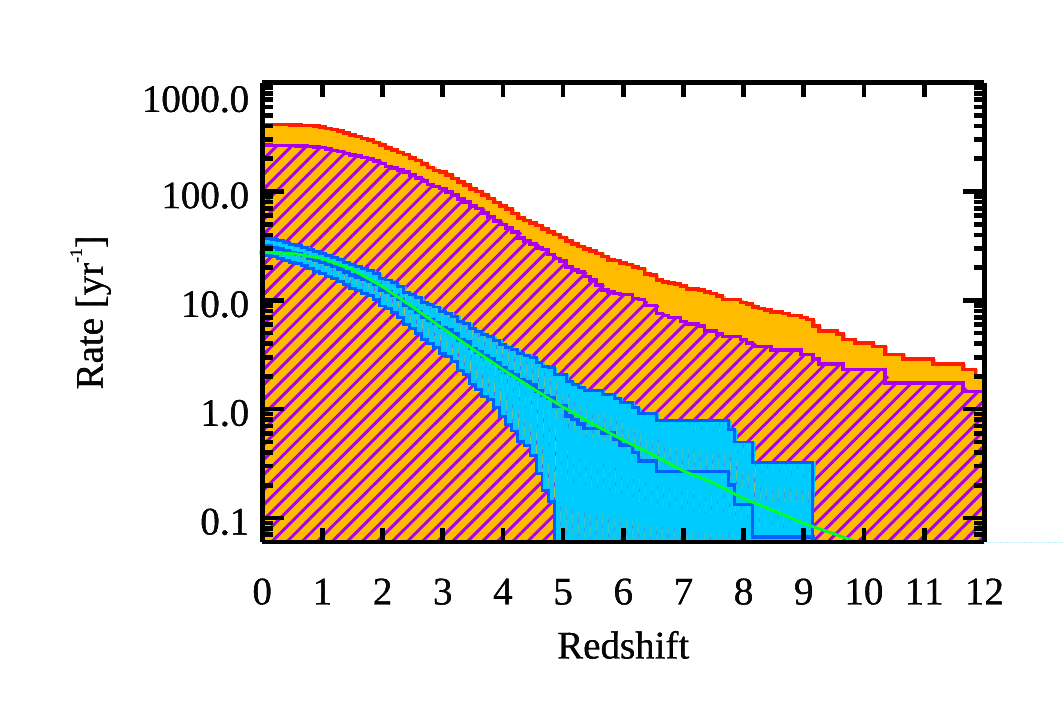}
\includegraphics[width=0.45\textwidth]{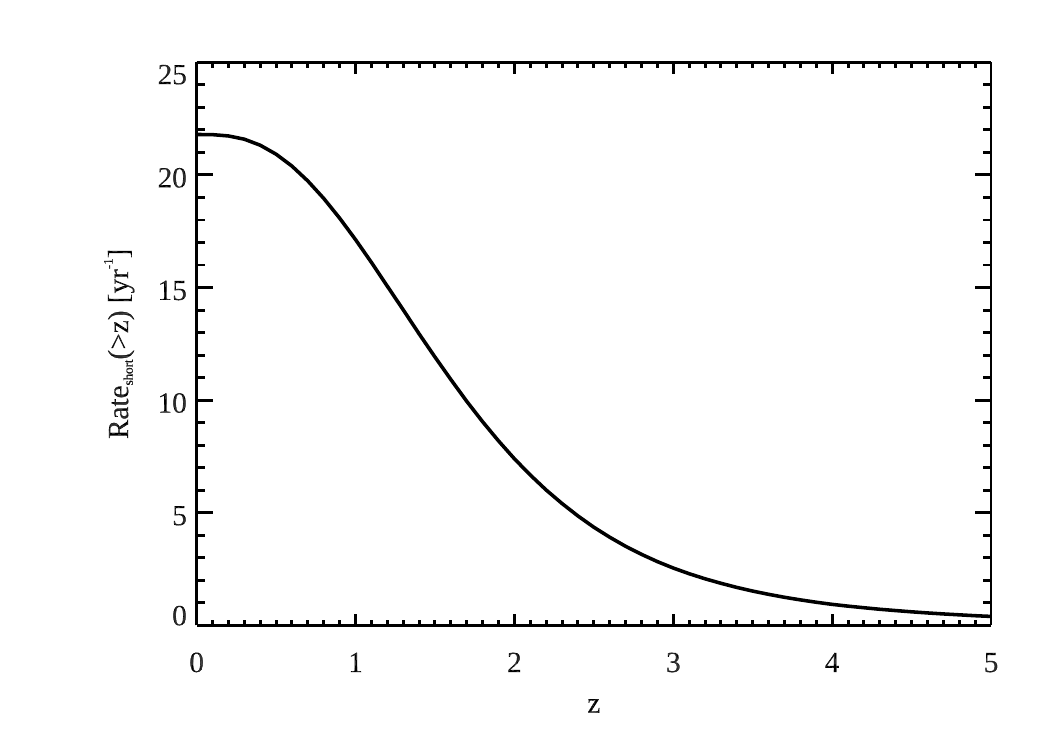}
\caption{
Left plot: 
{Observed GRBs with known redshifts measured in 2005--2020 (blue line and filled cyan area representing $1 \sigma$ uncertainty) superimposed on the anticipated frequency of detection of long GRBs by \textit{THESEUS} (orange histogram).}
The purple hatched histogram {shows the GRBs that are expected to have a redshift measured by either \textit{THESEUS} or ground-based facilities' telescopes}. 
{The model that fits the observed distribution used to make predictions for \textit{THESEUS} is represented by the green curve. }
\textit{THESEUS} {is expected } to detect {one to two} orders of magnitude more GRBs than Swift at any redshift, and most {importantly} in the high-redshift {range} ($z > 6$). 
Right plot:
Cumulative redshift distribution of short GRBs detectable with \textit{THESEUS}/XGIS per year of mission. 
{Theoretically}, short GRBs can be detected at high redshifts $z > 4$ with a rate of $\sim$1 event per year.
This figure is adopted from \cite{Ghirlanda2021}.
}
\label{fig:ghirlanda1}
\end{figure}

The left panel of Figure~\ref{fig:ghirlanda1} shows the expected detection rate of long GRBs by \textit{THESEUS} compared with observed GRBs.
{The orange histogram depicts the cumulative distribution of the GRBs detected by the SXI and/or XGIS (the bursts with measured $z$ are marked in purple), while the blue histogram presents the distribution of GRBs with known redshifts detected from 2005 to the end of 2020.
The distribution of the GRBs detected by \textit{THESEUS} was acquired based on the anticipated IRT capabilities and on the assumption of a ground follow-up rate of 50\% for the GRBs at $z<5$.
It is noticeable that \textit{THESEUS} is expected to detect an order of magnitude more bursts than \textit{Swift} does, especially in the high-redshift domain ($z>6$) \cite{Ghirlanda2021}.
It is expected that the redshifts for the majority of GRBs detected by \textit{THESEUS} will be measured (onboard or on the ground).}
{The cumulative distribution plotted in the right panel of Figure~\ref{fig:ghirlanda1} represents the annual detection rate of short GRBs by XGIS, not corrected for mission observation efficiency.
\textit{THESEUS} is supposed to acquire a statistically significant sample of short GRBs, including high-redshift ($z \lesssim$ 4--5) events. 
Considering that the distribution of the spectral lags of short GRBs and long GRBs differs significantly, advances in research on short GRBs are very important for such sophisticated studies of the QG effects as we discuss in this paper.}
{Thus, the sample of GRBs with measured redshifts obtained by THESEUS is the most promising for the application of the described technique to study LIV.}
\vspace{-10pt}

\section{Conclusions}
\label{sec:conclusion}
Various QG theories predict LIV, which can manifest itself as the dispersion of the speed of light.
The method that we propose {to disentangle and constrain} the QG delays from the intrinsic {spectral lags} in GRB light curves is based on the assumption of the constancy of the rest-frame intrinsic spectral lags and on the linear dependence of the GRB spectral lag on both the photon energy and function of the GRB redshift.
The ability to collect a large sample of GRBs with known redshifts is crucial for this type of study, as the precision of the QG effect measurement can be improved by expanding the data set.
Currently, redshifts are measured spectroscopically or photometrically for $\lesssim 500$ GRBs.
Thus, indirect estimates of the redshifts from the prompt emission observables are necessary to obtain a large GRB sample for LIV studies using the commissioned instruments.
The sample of GRBs collected by \textit{Fermi}/GBM could provide a promising opportunity to apply the aforementioned technique, thanks to its extensive trigger statistics ($\gtrsim$3500 GRBs up to date) and sophisticated data acquired with a high temporal and spectral resolution, which could allow precise measurements of the rest-frame spectral lags.  

The \textit{THESEUS} mission is likely to initiate a breakthrough in this field of fundamental physics as, thanks to the combination of its unique characteristics, the observatory will collect one order of magnitude more samples of GRBs with known redshifts than are currently available, which will not only allow the use of GRBs as cosmological tools but will also enable us to constrain the QG theories.
Moreover, thanks to its capability to detect the GRB emission in the relatively soft energy band of 0.3--5~keV, \textit{THESEUS} could provide a unique opportunity  not only to constrain the empirical and physical GRB models but also to expand the data range, providing more accurate constraints on the QG energy from the lag-energy plane.
Due to its high sensitivity, \textit{THESEUS} will also allow advances in the study of short GRBs; in particular, XGIS will be able to detect short GRBs up to z$\sim$4--5, which is important as the spectral lags of short GRBs essentially differ from the ones of long bursts.
Thus, the \textit{THESEUS} mission could make a significant contribution to the study of the QG effects using GRBs.

We thank the anonymous referees for helpful comments on the manuscript.
Some of the authors wish to thank ASI and INAF (agreements ASI-UNI-Ca 2016- 13-U.O and ASI- INAF 2018-10-hh.0), the Italian Ministry of Education, University and Research (MIUR), Italy (HERMES-TP project) and the EU (HERMES-SP Horizon 2020 Research and Innovation Project under grant agreement 821896) for the financial support within the HERMES project.
L. B., A. S. and A. T. acknowledge funding from the Italian Ministry of University and Research (MUR), PRIN 2017 (prot. 20179ZF5KS), “The new frontier of multi-messenger astrophysics: follow-up of electromagnetic transient counterparts of gravitational wave sources” (PI: E. Capellaro).
T.d.S., L.B. and A.S. also acknowledge the financial support of PRIN-INAF 2019 within the project ``Probing the geometry of accretion: from theory to observations'' (PI: Belloni).

\clearpage
\bibliography{quantum_gravity}{}
\bibliographystyle{aasjournal}

\end{document}